\documentclass[prd,aps,preprintnumbers,amssymb]{revtex4}
\usepackage{graphicx}% Include figure files
\usepackage{dcolumn}% Align table columns on decimal point
\usepackage{bm}% bold math
\begin{document}
\title{
%\hfill\vbox{\hbox{\small PP-04- \qquad\qquad}}\\
%\hfill\vbox{\hbox{\small SU 4252-804 \qquad\qquad}}\\
BETTER  PHYSICS TEACHING CAN INCREASE PHYSICS ENROLLMENT
}
 \author{Samina S. {\sc Masood}}
 \email{masood@uhcl.edu}
 \affiliation{Department of Physics, University of
Houston Clear Lake, Houston TX 77058}
\begin{abstract}

Our main goal is to develop plans to increase physics enrollment. Once again
we thoroughly analyze the problem from the beginning and reach the
conclusion that the most appropriate starting point in this direction should
be to look into K-12 teaching. We give a few recommendations to improve
science/physics teaching at K-12 level. It is proposed that the quickest way
to make some advancement is to start teacher training or refresher courses
for school teachers to fill up their gaps in knowledge. We suggest a
comparison of the affectivity of different methods of teaching to decide
which one of them works better under what type of circumstances. We also
propose a few steps to improve physics teaching standards at the higher
levels.
\end{abstract}
\maketitle

\section{WHY  DO  WE  NEED  PHYSICS?}

\smallskip

We start our analysis by reminding ourselves with the future needs of
societies. In this technological era only those nations that can keep a
balance between the development in science education and the development in
technology will rule the world. Keeping that balance is perhaps the only way
to get a respectable position in the technology based future and to enjoy
the modern time lifestyle. Therefore, mankind will have to develop
scientific approach towards life through the basic science education.
However the standard of basic science level is also being changed with time.
When we talk about science education, we mean all kind of sciences but
talking about technology, we have to give physics a little more importance
than some of the other branches of sciences as it is helpful to understand
almost each and every branch of science, especially in relation to
technology. Physics obviously has many more applications to technology and
is actually a language of technology in modern era. It should, therefore be
realized that learning basic physics for every citizen is almost as
important as learning a common language to completely share the joy of a
social life. In other words physics is the mother of engineering and
technology. It has suffered most because it is always associated with
mathematics and is an accepted difficult subject. It could easily be
decoupled from mathematics at the elementary level. The interested students
who are motivated enough to learn the subject in detail, would be able to
learn required mathematics at a later stage. However, to really improve the
interest in science education, because of their interest, we need to improve
mathematics teaching right from the very beginning. This will not only help
to develop a well-planned and financially organized society but also help to
increase the physics enrollment as well as the quality of physics education.

\smallskip

If we want to do anything to improve the physics enrollment and learning, we
have to improve physics teaching. Unfortunately, physics teaching is one of
the most ignored areas in several ways. We have very effective and dedicated
teachers, very learned physicists and even enough motivated students. The
problem lies in the system itself and we have to look at the origin of this
problem. We are convinced that the roots of this issue arise from public
schools. At K-12 level we do not have enough physics majors to introduce
physics in an appropriate way to the young children. When we go back to look
in to the reasons behind it, we find that a few years ago, a physics degree
was considered to be a paying degree. Even then, a physics degree holder
would not prefer to teach as it is not a well-paying job. It is still not a
well-paying job but the general cuts in physics jobs could have attracted a
bigger ratio of physics majors to teaching. But we could not still improve
the ratio of teachers with a physics degree. As a consequence, we are losing
interest in physics among children and are expected to be even unable to get
the minimum required American national scientists. On the other hand,
because of the changing situation with the immigration rules, it might even
not be possible to bring as many physicists from other countries, especially
for some special projects running in sensitive areas. We still have
students, but as soon as they get the degree, they go back to their
countries. On the contrary, American students are losing interest in
physics. We need to plan now to avoid that situation where we may not be
able to win the complete supremacy in technology, especially to compete the
continuously growing trend almost everywhere else on the globe. The
US-Government has to handle these long-term issues more seriously.

\smallskip

Most of the physicists after obtaining Ph.D. prefer to join 
computer-related  jobs
in industry or sometimes even provide cheaper labor by teaching as adjuncts
in colleges. Some of them are undoubtedly great teachers or researchers and
hiring them by colleges may not be any financial burden as they may attract
a large number of students and can bring enough research money to compensate
their salaries or even increase financial resources for their institutes.
This is a well-known fact that even though the science is not being promoted
in many institutes; most of the external funds are still generated by
science faculty for all schools. However, teaching positions have badly been
reduced in science because of the general trend of hiring adjuncts more than
the actual faculty. We have discussed the problems related to hiring
adjuncts in detail in Ref. [1]. A typical ratio of adjuncts to full-time
faculty is accepted to be 35\% or less, and it has now risen much more than
that especially in state colleges and two year colleges. In state colleges
 this ratio is sometimes reversed. Several state colleges have just one
full-time physics professor and even run the graduate program. Adjuncts
salaries could be almost half of the full time faculty members but the 
resources
which full-time faculty can bring could be at least double the salary of
adjuncts. It is not difficult to calculate that the colleges are in fact
losing money and resources by hiring adjuncts and ultimately end-up closing
some of the science programs without looking at the cause of decrease in
enrollment. This is especially common in two-year and four year colleges. We
have seen several state colleges where physics programs are being run by a
single full-time faculty member or even by non-physics faculty. This badly
affects all the related programs and spoils the quality of physics teaching
in such places. The cutting budgets of colleges have raised this trend to
hire less expansive teachers, which in turn affects the standard of
education and the decrease in science enrollment, in general. Physics has
suffered greatly by this policy. The impact of this happening is clearly
seen from the data of physics enrollment during last few years. There could
sometimes be an increase at the undergraduate level at some places [2]
because more programs have started requiring physics courses. However, we
are addressing here the enrollment in physics programs itself instead of the
elementary level physics courses required by several other majors.

\smallskip

When the question of physics education comes under discussion, we have to
differentiate between science majors and non-science majors. We all agree
that physics is compulsory for all engineers and science majors. They can
excel in their careers with a better physics background only. Actually at
the research level, physicists can even work as engineers. So there are
always physics courses for these programs and they are almost standard.
There is always room for improvements, and we in the University of Houston
Clear Lake are trying to improve our undergraduate physics teaching to a
greater extent to be able to attract more students in graduate program. We
could see the immediate effect through the increase in enrollment by 20\% or
more. This increase is with respect to the physics enrollment in subsequent
years. However, the ratio of physics majors with majors in other deciplines
is not increasing so much.

\smallskip

However we want to step back a bit and see how we can get more students in
science in general and physics in particular. Our main concern is to
increase physics enrollment and attract more interest in the subject itself.
Science and engineering majors have to take physics to satisfy the
requirement of their programs. Whereas non-science majors usually put
physics at the bottom of the list of courses of their choice among the
science courses. Or we can safely say that there would be hardly anybody who
would like to take physics if he or she does not have to take it for the
program requirement. Our main goal is to bring physics at the equal level
with other introductory science courses, although it is always considered to
be a difficult subject. A better mathematics background definitely makes a
difference and is proven to be helpful always. This is the actual situation
in some of the other countries which have not yet adapted to technological
development so much.\smallskip

\smallskip

Countries like India, China and Japan are working hard to teach mathematics
and science to their children and we can see the speed of their development
regardless of their limited resources. But we are perhaps too busy in
enjoying the present and may even ignore our future. Falling behind in that,
we forgot to do much for securing our future.

\smallskip

Now we have reached a point where we always hear, ` I hate math'. `Physics:
it is hard and boring' or `physics is so difficult that I cannot
understand.' `What is the use of studying physics, it has no practical
applications. We cannot even earn enough money with that' and most tragic
statement which we hear is, `physics is interesting but I cannot starve' `I
would enroll in physics if someone promises to pay me in this career like
others.' Moreover, students at the higher level usually say math is boring
and physics is math. All the negative comments about math are associated to
physics along with its own image of being difficult and boring subject.

\smallskip

Programs such as `No Child Left behind` [3] should include science along
with mathematics. Every child deserves to be exposed to science in relation
to daily life. We need to help our children to learn how to contribute to
assure a safe and healthy future. We need to make them aware of our future
needs. Physics is an amazingly important subject in this regard. It is
considered the mother of technology. Especially in this technological era,
it helps to enjoy the technological development in its entirety. All
branches of engineering have their roots in fundamental physics. Knowing
this, we need to teach physics adequately at all levels and maximum students
should be attracted to physics to continue leading the entire world in
technology. Unfortunately, we are not doing very well there. Physics has
been tied up with mathematics to the extent that it is usually taken for
granted that you can only understand physics if you are exceptionally smart
in mathematics. It is not incorrect that physics could properly be
understood with the help of mathematics only and it could be fully described
mathematically. Still it is not all mathematics. It is only one of the
applications of mathematics. It has experimental and observational
components built-in there. No physics course can be completed without the
integral component of laboratory work. We somehow completely ignore the
applied physics which has much more overlap with engineering itself. On the
other hand math as a subject is considered to be a subject for very smart
students. In this situation if we need to attract more students in science
we have to start attracting students in science right from the beginning.
So, the teachers from elementary, middle or high schools can play the most
important role in developing an interest in science. For this purpose, our
school teachers need to be trained in physics so that those who were not
physics majors but are teaching physics can have adequate knowledge to
satisfy immediate needs of the corresponding subject. Moreover, teachers are
needed to be aware of the latest development in science and should be able
to share the fascination of new discoveries with children who are our future
scientists. In this situation if at the early stages of school, students can
get some inspiration about science subjects from their parents or teachers;
getting into these fields is no more difficult for them. This is what is
really happening in most of the countries in Asia, Russia and Japan.

\smallskip

\smallskip

In this era of science and technology where computers have made it possible
to visualize things a lot through simulation, we do not want to learn
through equations or our imaginations only. However, the beginners need
motivation by the fascination of physics through visual mode of learning. It
even compels those beginners to learn whatever is involved in the learning
of technical language of the subject such as mathematical equations and
computation. Mathematics itself is not interesting to most people. It is
a great tool and is learned when you want to understand some other concepts
and ideas in business, science or computation, etc. If you can get adequate
mathematics training for business majors why cannot we do the similar amount
of mathematics training for physics majors? The courses like methods of
mathematical physics can be taught very well in a physics department to
satisfy the immediate needs of physics majors in the beginning. I guess such
courses are much more helpful when they are taught in the physics department
and students do not have to go with several courses in mathematics which are
sometimes not offered in a convenient time for physics students. Either the
mathematics and the physics faculty has to work together to offer
appropriate mathematics courses for physics students at the time of their
convenience or some mathematics courses are especially designed for physics
students and are taught as a part of physics program.

\smallskip

In this situation I guess physics teaching needs a lot of improvement. Not
because the other science subjects are not equally important, just because
physics is equally important and still undermined for several reasons. One
of the big reasons is that the most of the science teachers in high schools
are science majors, but most of them are not the physics majors because we
do not have such a high enrollment in physics degree as we have in other
science subjects. This obviously affects the ratio of physics majors in the
recruited science teachers. The simple reason is again because it is
considered to be very tough to have too much mathematics and not to have so
much practical applications to life. There are several reasons behind it and
we have discussed them in Ref.1. How can we expect those teachers to make
physics attractive for students who did not choose to study it themselves?
At this point we just need to know how to come out of this mutually
entangled problem where the decrease in enrollment keeps on growing without
even giving a starting point or an end point. We just want to see how to
overcome this problem here. Let us start from trying to find the root cause
of this issue. I think it lies somewhere in school education where the
students develop a mathematical background and are exposed to science
subjects for the first time in life. We need to look at this carefully and
take some concrete steps to improve the level of science teaching in
schools. According to our analysis, the best starting point would be middle
and then high schools, where students are exposed to science for the first
time as a real subject. In good school districts we have a better ratio of
science teachers with physics degree. However, it is a big concern for the
educated community that the middle school teachers who almost give the first
proper introduction of science to students may not even have their major in
science subjects and if they do, may not be equally trained to teach the
subject, they are teaching. It has even be noticed in the elementary 
schools that sometimes  teachers may not be able to answer science 
questions appropriately. With a small number of physics majors, in
general, there is a big possibility that physics is being taught by those
people who may have just taken one physics course and have hardly passed it.
Should we expect from such teachers to attract our new generations to
physics and be able to prepare more physics majors. May be not! They can 
only promote subjects of their own
interest that they know better than others. This is quite well-known and
pretty understandable also that the teachers can always teach those 
subjects better than others which they like most and enjoy to teach. 
Moreover, those teachers prove to be better teachers for whom teaching is a 
passion 
and they love the subject itself. 

\smallskip

\section{IDENTIFICATION OF THE PROBLEM}

\smallskip

When we started investigating the root cause of this problem, everything
appeared to be tied up in a loop and we had to decide on a starting point by
carefully analyzing the situation in detail. The science community has a
special concern about the decreasing enrollment in science, in general, and
physics, in particular. This decrease in physics enrollment obviously can
lead to a situation in which we do not have enough physics majors to fill
the basic needs of employers or to introduce this subject adequately to our
children. As a consequence we will not be able to attract our children into
the subject. This is unfortunately what is happening in several school
districts nationwide. We have analyzed some data in order to look at this
situation quantitatively.

\smallskip

\begin{itemize}
\item  The nation's schools employ 21,000 physics teachers [4]. Two-thirds
of them do not have a degree in physics or physics education. The most
common degree in science is biology, and in this degree only two semesters
of general physics are required. Students, who are taught by biology majors,
usually do not do as well in physics. Thus there is a great need for more
physics teachers and for preparation of cross-over in-service teachers. In
this situation these out-of-field teachers should required doing some
preparatory courses in physics before they start teaching physics. It will
be a much faster approach than bringing in all new physics majors to teach
physics.
\end{itemize}

\smallskip

\begin{itemize}
\item  Senior teachers with physics or physics education degrees may have
studied physics many years ago. They may be very good teachers and are
perhaps the only ones who are playing a major role in maintaining our
physics enrollment. However, these teachers will be better prepared to face
their professional challenges in attracting students to physics if they are
aware of new developments in the subject and if they are exposed to new
teaching techniques.
\end{itemize}

\smallskip

\begin{itemize}
\item  Moreover, it has been noticed that in most of the teacher's
certification programs the minimum requirement to get into the program is
the same as the minimum requirement of getting in the school [5]. Moreover,
this certification just requires passing grades (C or above) in any subject.
This standard is not attracting our best physics students to a teaching
career as a majority of the students with good grades can get into other
higher-paying careers. Those who could not go to another career would go
into teaching as an alternative. This obviously affects those subjects which
need to be taught by people who are really interested and knowledgeable in
the subject itself and can teach it adequately to attract children to the
discipline.
\end{itemize}

\smallskip

\begin{itemize}
\item  We have looked at the salary structure of schools in Texas [6]. The
salary difference between a bachelor and master degree is just \$1,000
dollars per annum and with Ph.D. teachers only get \$200 extra in a year as
compared to the master degree. In this situation, why would a teacher spend
the extra time for a Ph.D. or even a master? So, there are no real
incentives for teachers to go for higher education.
\end{itemize}

\smallskip

\begin{itemize}
\item  On the other hand, there are several Ph.D.'s available who could have
taught in schools, but can not as they do not have the school certification.
They do not want to spend money and time to get certification and then end
up getting low school teacher salary without getting any benefit of higher
specialized degree. They instead accept computer or other related
engineering jobs in industry and pretty much leave physics and get much
higher salaries as compared to school teachers.
\end{itemize}

\smallskip

\begin{itemize}
\item  Recruiting students in physics program is also a very tricky job.
Physics enrollment is already significantly below the desired level. There
are not many people taking interest to join physics. In this situation we
have to relax enrollment criteria a little bit to keep the enrollment above
the minimum requirement for any program.Sometimes, because of the decreasing
interest in mathematics, we have to compromise on the mathematics
background. On the other hand, it is noticed that those students who do not
have ever taken physics but have better background in mathematics could
easily manage to pick-up the calculus-based physics course. These students
may do much better than those who have even taken algebra based physics but
have not completed the mathematics courses. In this situation we may have to
think about revising our prerequisites. If we cannot fix the problems
related to mathematics teaching in schools, we may want to fix it at the
college level to get help in developing a better physics background.
However, in this attempt of increasing standards, there is a danger of
students dropping out the existing enrollment level due to a greater failure
rate also. Therefore, just to keep program running, sometimes we have to
compromise on those things which should not have.
\end{itemize}

\smallskip

\begin{itemize}
\item  People who are working in physics education are developing teaching
models which work in their setup; however, it may or sometimes may not be so
effective in other situations. Some of the good examples are modeling [7-10]
techniques and thinking skills [11]. John Clement[12] has given a comparison
of gains between FMCE [13] and some of the other methods[14,15].
\end{itemize}

\smallskip

\section{ROOTS OF THE PROBLEM}

\smallskip

We have just mentioned a few reasons for the decreasing interest in physics.
We could analyze the situation through our personal experience, some data
analysis and the exposure to the existing literature. But this analysis
cannot be a thorough analysis as it has too many variables. All our results
somehow seem to depend a lot on local circumstances and individual factors.
We can pretty much find the reasons of whatever is happening and how our
teaching becomes a sensitive function of several variables. Some of them
have already been discussed in Ref. [1]. We will postpone the detailed
analysis of these issues and will concentrate on finding a solution. We can
even list some of the well-known variables as follows:

\smallskip

\begin{itemize}

\item  instructional resources

\item  educational training

\item  financial satisfaction

\item  professional commitment

\item  career goals

\item  personal priorities

\item  academic interests

\item  Family background.

\item  Work atmosphere

\end{itemize}

\smallskip

Teaching can be made a lot more effective using the modern technology.
Computer simulations make it possible to visualize things that could not
have been imagined only a few years ago. Because of this, we do not want to
learn through equations or our thinking skills and visualization only.
Beginners need motivation via the fascination of physics through a visual
mode of learning. Demonstrations, experiments and simulations play an
important role in this fascination. It even compels them to learn whatever
is involved in the learning of the technical language of the subject,
whenever it is needed.

\smallskip

Modeling [7-10] and other visual methods are definitely very effective and
in some cases, they are the only methods to teach some of the students.
However, we should be able to train at least some of the students in a way
that they can look at the situation intuitively and able to make an educated
guess before even doing any calculations. Thinking skill is a key to
research contribution. For example, mathematics itself is never interesting
for most people. It is learned when someone wants to understand the basic
concepts and ideas in business, science, computation, or some other fields.
If we can deliver adequate mathematics training for business majors why
can't we teach our students just as much mathematics in a physics class?

\smallskip

I guess all of us have noticed that a better mathematics background leads to
the development of a better ability of visualization of the physical
concepts. In this situation, we should try to segregate between those
students who want to go for a physics program and the ones who just want to
take physics for other programs. Even elementary physics courses like
calculus-based physics should have at least two different groups. Having
these two groups is all possible, but we do not want to hire enough teachers
and compromise our standards and even our future. It is a real
disappointment when a teacher of calculus-based physics course has to teach
two hundred students in the same class at a lower level to make it
understandable for majority of the class. But, doing that, we may discourage
or lose some of those twenty students who may want to become future research
physicists and cannot comprehend enough in this type of class setup. On the
other hand, it is almost impossible for students to get reasonable
individual help from the instructor. Failure rate in these big classes is
usually higher and students are unhappy as they cannot communicate with the
instructor directly. Usually teaching assistants, who are graduate students
themselves, help students.

\smallskip

Moreover, there is a well-known hierarchy in teaching standards between
community colleges, state colleges and the research universities. We need to
start some work from the school level. Research universities try to keep
standards at the cost of high failure rate. Students can save money in the
same order and compromise on the standards in the reverse order. Partnership
between community colleges, state colleges and research universities is also
a good way to overcome the problems due to this hierarchy.

\smallskip

Before we discuss about how the physics enrollment can be increased in the
presence of all the above mentioned factors, we want to indicate that the
causes of problem can be found at each and every institutional level
including middle schools, high schools, community colleges, state colleges
and research universities. However, a few things can still be recommended at
all different levels and should be applied simultaneously to overcome these
issues.

\smallskip

\section{HOW TO IMPROVE THE SITUATION-a few recommendations}

\smallskip

We propose a few recommendations to start efforts at several institutional
levels at the same time. We want to discuss them one by one and see what can
be done about it. We start from lower to higher level.

\smallskip

\begin{itemize}
\item  At K-12 level the problem are briefly indicated. In the long run we
should plan to hire physics majors to teach physics at school levels.
However, as an immediate solution, we would recommend starting summer
workshops for school teachers or summer refresher courses for them. These
courses should be organized by the physics department or program of a school
and are taught by the physicists. All the physics teachers should be
required to these courses at least once in three years. These courses are
especially needed for those teachers who are teaching physics based middle
school courses and in some cases, high school physics courses and have not
even opted to be a physics major themself. We want to initially collect data
on the ratio of non-physics majors to physics majors among middle school and
high school physics teachers in several school districts and then would like
to investigate the functional dependence of science majors on the
qualification of teachers in general. For this purpose we will have to
collect data on the number of students going for science majors from our
test school districts. We would also like to see how our program makes a
difference in this ratio of students going for science majors.
\end{itemize}

\smallskip

\begin{itemize}
\item  We are also interested to keep a continuous check on our program. For
that purpose we want to work very closely with the school of education and
keep a continuous check on the affectivity of our program. From the school
of education, we can share expertise in data collection and sometimes even
research students who are interested in specializing in science education
along with introducing research in the topics of physics and science
education in the college of science and computer engineering. This way we
should be able to generate some inter-college collaborative work within the
campus. In the first year we need to collect data on all the middle school
teachers from almost all school districts of Houston region and then will
compare it with the existing school district ratings. We will also gather
data about the ratio of students going to colleges and ratio of science
majors getting out of each school. Then after introducing our program, we
would like to see the impact of this program on these ratios. From here we
can not only apply this program to schools in Texas, we can even expand it
nation-wide and who knows how much success can be achieved!!
\end{itemize}

\smallskip

\begin{itemize}
\item  Some of the school districts have special programs like
Science-Magnet program [16], Science-Olympiad [17] etc. These programs are
very effective; however a partnership between schools and colleges could
make the program a lot more effective. If these schools can arrange some
science workshops for students at individual grade level or arrange science
fairs, that will be very helpful. Such programs will at least increase the
ratio of science majors by 10-20\%. This number is based on the data
collected from overseas, where the circumstances may be different from here.
Though, we still believe that it will not be very different over here.
\end{itemize}

\smallskip

\begin{itemize}
\item  In the community colleges or some four year colleges which have
physics program to support other degrees, the standard is decreased when we
see a master-degree holder engineer as the program coordinator for physics.
Just to keep his or her position secured, he or she will never prefer to
hire a physics Ph.D. as a second or third faculty member. These
qualification deficiencies may some times lead to generate inappropriate
relationship between different parties and spoils the academic atmosphere.
How can we axpect a good physics program if the program coordintor has a
limited knowledge in the subject. This is how the standard of program is
severely damaged, sometimes. So we have to take care of this problem. Even a
community college or a two- year physics program should have at least one
physics Ph.D. as the program chair. That will make a difference.

\item  We need to reduce the difference of standards between same types of
courses. One example is an algebra-based physics as the most common course
which is taught as AP (Advanced Placement) physics course in high schools,
college physics course or algebra-based physics course in community colleges
and state colleges. This is the type of problem with calculus-based
university physics course. This course has a large difference of standard
between state colleges and research universities. Some times physics
programs make a decision in accepting credits from other colleges to keep
the same standards. However. students are discouraged or cannot afford time
and money to go to join physics program and take those courses again, which
they have already done in other colleges
\end{itemize}

\smallskip

\begin{itemize}
\item  Money and time are other big concerns for students who join a physics
program. We in the University of Houston Clear Lake have introduced a
non-traditional physics degree program [18] where we have a majority of
students as in-service part time students. These students are not full time
students and obviously have some limitations of time due to their work
responsibility. However, it has been noticed that these students are still
more motivated students as they choose to learn physics and have job
satisfaction and are satisfied financial issues. These types of programs are
very good to promote physics and increase physics enrollment as these
students come from the background where they do not worry so much about
finding a job after getting physics degree and also have no rush to get a
degree. They take one or two courses every semester and study it seriously.
Good grades are also a kind of requirement for such students as they can
only get financial support from there companies for studies if they get at
least B-grade. Our class size for graduate course is usually 8 to 10 which
is comparable with many research universities. And students are doing even
better than research universities, as they have lesser course load and can
manage it easily with their work. These types of programs could be developed
in other state colleges and universities, especially in the campuses which
are situated on favorable locations such as UHCL.
\end{itemize}

\smallskip

These types of programs could be started at different levels simultaneously
and the immediate increase in the standard of physics education as well as
the physics enrollment. We are expecting the most immediate dissemination to
the students of the teachers who attend this workshop. Students will have
the most effective modes of instruction and will be better prepared in
physics. They will learn better how to understand physics concepts, how to
solve problems, how to use computers as scientific tools, how to
communicate, how to transfer their knowledge to daily life and how to become
lifelong learners. We intend to present workshops on these topics at
regional meetings and conferences of the American Association of Physics
Teachers (and/or the State Science Teachers' Association).

\smallskip

These types of programs may initiate collaborative efforts between Schools
of Education and Schools of Science in several campuses, which will set-up a
trend for further inter-school programs and help to develop
inter-disciplinary research trends in US campuses in general. We are working
on developing inter-disciplinary collaborations at UHCL.

\smallskip

This partnership of campuses to public schools will automatically introduce
the ongoing science program in different campuses and students will easily
be able to decide on their choices. It is also noticed that if kids get
opportunity to directly talk to highly educated professional community, they
may be impressed by their personalities. Due to that reason sometimes they
idealize them and decide to follow their educational path. This way, we can
attract more students in science. This will lead to the growth in campus
enrollment in general and will help to grow the science enrollment in
particular. The growing trend of science growth, in the presence of our
geographical location, will help us to develop stronger ties between
different educational institutions. It s expected to be a very effective 
way to attract more students to science subjects and prepare for our 
future properly. 

\smallskip

\section{References and Footnotes:}

\smallskip

\begin{enumerate}
\item  Samina S. Masood, ` The Decrease in Physics Enrollment'
physics/0509206

\item  http://www.eurekalert.org/pub\_releases/2007-01/aiop-hsp011007.php

\item  https://www.k12.wa.us/ESEA/pubdocs/FAQSupplementVersesSupplant.pdf

\item  Ms. Debra Hill from Friendswood Independent School District in
Houston area and several other websites.

\item  We have checked with several schools offering school certification
about their requirements and do not want to mention all the names here.

\item  We have looked at the salary announcements of several Houston Area
Schools in this regard.

\item  Malcolm Wells, David Hestenes and Gregg Swackhamer, `A Modeling
Method: for high school physics instruction' Am. J.Phys.63(7),606(1995)

\item  David Hestenes and Jane Jackson, ''Partnerships for Physics Teaching
Reform-a crucial role for universities and colleges.'' And several programs
from University of Arizona.

\item  David Hestenes, `Modeling Methodology for Physics Teachers'
Proceedings of International Conference on Undergraduate Physics Education
(College Park, August 1996)

\item  Jane Jackson,
http://modeling.asu.edu/R\&E/ModelingWorkshopFindings.pdf

\item  Anton Lawson' Book entitled, `Science Teaching and the Development of
Thinking'

\item  John Clement, `Physics Education Research: Are your students learning
what you think?' A presentation in University of Houston Clear Lake (Jan.
2007).

\item  Details can be found in several places, see for
example:http://perlnet.umephy.maine.edu/materials/index.html.

\item  David R Sokoloff and R.K.Thornton, The Physics Teachers, Vol.35,
340(1997) and several other references.

\item  Lillian McDermott, Book entitled, `Physics by Inquiry.'; Lillian
McDermott, Am. J.Phys. 59,301(1991) and several other references;

\item  Clear Creek ISD in Houston area in Texas

\item  Fayetteville-Manlius High School in New York and several other school
districts (nation-wide) have this program.

\item  Details of this program can be found in David Garrison, 'Development
of a Comprehensive Physics Program at a Non-traditional Upper-level
Undergraduate and Graduate Small University' APS Forum on Education Spring
2006 Newsletter.
\end{enumerate}

\end{document}